\documentclass[12pt]{article}
\usepackage{latexsym}
\usepackage{amsmath,,calrsfs}
\usepackage{amsfonts}
\usepackage{amssymb}
\usepackage{amscd}
\usepackage{bbm}
\usepackage{fancybox}
\usepackage{cite}
\usepackage{amsmath,amsfonts,amsbsy}
\usepackage{pstricks,pst-node}
\usepackage[small,bf,hang]{caption2}
\usepackage{graphicx}
\usepackage{epsfig}
\usepackage{psfrag}
\usepackage{comment}
\usepackage{hyperref}

\usepackage{float}

\psset{unit=1.3cm,linewidth=.5pt,radius=.2}  

\usepackage{multirow}                     
\usepackage{float}                          
\usepackage{lscape}                         
\usepackage{bm}

\newcommand{\cL}{{\cal L}}

\newcommand{\bb}{\bar\beta}

\newcommand{\beq}{\begin{equation}}
\newcommand{\eeq}{\end{equation}}
\newcommand{\bi}{\begin{itemize}}
\newcommand{\ei}{\end{itemize}}
\newcommand{\bt}{\begin{tabular}}
\newcommand{\et}{\end{tabular}}
\newcommand{\bc}{\begin{center}}
\newcommand{\ec}{\end{center}}

\newcommand{\be}{\begin{equation}}
\newcommand{\ee}{\end{equation}}
\newcommand{\bea}{\begin{eqnarray}}
\newcommand{\eea}{\end{eqnarray}}
\newcommand{\ba}{\begin{array}}
\newcommand{\ea}{\end{array}}

\def\bbox{{\,\lower0.9pt\vbox{\hrule \hbox{\vrule height 0.2 cm
\hskip 0.2 cm \vrule height 0.2 cm}\hrule}\,}}
\newcommand{\dsl}{\pa \kern-0.5em /}

\font\mybb=msbm10 at 12pt
\def\bb#1{\hbox{\mybb#1}}

\def\bR {\bb{R}}

\makeatletter \@addtoreset{equation}{section} \makeatother

\def\slashchar#1{\setbox0=\hbox{$#1$}           
   \dimen0=\wd0                                 
   \setbox1=\hbox{/} \dimen1=\wd1               
   \ifdim\dimen0>\dimen1                        
      \rlap{\hbox to \dimen0{\hfil/\hfil}}      
      #1                                        
   \else                                        
      \rlap{\hbox to \dimen1{\hfil$#1$\hfil}}   
      /                                         
   \fi}

\begin{document}

\begin{titlepage}
\begin{center}

\vskip 1.5cm

{\Large \bf On asymptotic charges in 3D  gravity}

\vskip 1cm

{\bf Eric A. Bergshoeff\,${}^1$,  Wout Merbis\,${}^2$ and
Paul K.~Townsend\,${}^3$} \\

\vskip 25pt

{\em $^1$  \hskip -.1truecm
\em Van Swinderen Institute, University of Groningen, \\ Nijenborgh 4,
9747 AG Groningen, The Netherlands
 \vskip 5pt }

{email: {\tt  E.A.Bergshoeff@rug.nl}} \\

\vskip .4truecm

{\em $^2$ \hskip -.1truecm
\em  Universit\'e Libre de Bruxelles and International Solvay Institutes,\\ Physique Th\'eorique et Math\'ematique,
Campus Plaine - CP 231, \\
B-1050 Bruxelles, Belgium \vskip 5pt }

{email: {\tt  wmerbis@ulb.ac.be}} \\

\vskip .4truecm

{\em $^3$ \hskip -.1truecm
\em  Department of Applied Mathematics and Theoretical Physics,\\ Centre for Mathematical Sciences, University of Cambridge,\\
Wilberforce Road, Cambridge, CB3 0WA, U.K.\vskip 5pt }

{email: {\tt P.K.Townsend@damtp.cam.ac.uk}} \\

\end{center}

\vskip 0.5cm

\begin{center} {\bf ABSTRACT}\\[3ex]
\end{center}

A variant of the ADT  method for the determination of  gravitational charges as integrals at infinity 
is applied  to ``Chern-Simons-like'' theories of 3D gravity, and the result is used to find the mass and angular momentum 
of the BTZ black hole  considered as a solution of a variety of massive 3D gravity field equations. The results agree with many 
obtained previously by other methods, including our own results for ``Minimal Massive Gravity'', but they disagree
with others, including recently reported results for ``Exotic Massive Gravity''.  We also find the central charges of the 
asymptotic conformal symmetry algebra for the generic 3D gravity model with AdS vacuum and discuss implications for 
black hole thermodynamics.

\end{titlepage}

\newpage
\setcounter{page}{1}
\tableofcontents


\section{Introduction}
\newcommand{\m}{\mathfrak{m}}
\renewcommand{\j}{\mathfrak{j}}

In General Relativity (GR), conserved `charges' such as mass and angular momentum are generically expressible only as integrals
at spatial infinity; the prototype is the Arnowitt-Deser-Misner (ADM) mass formula for an asymptotically-flat spacetime \cite{Arnowitt:1962hi}. The method used to 
derive this  formula can be adapted to spacetimes that are  asymptotic to other ``background'' solutions of Einstein's field equation for which there
is a spatial infinity,  as shown by  Abbott and Deser who focused on the anti-de Sitter (AdS) case \cite{Abbott:1981ff}; the scope of the method was later extended by Deser and Tekin \cite{Deser:2002jk}.  In principle,  this Abbott-Deser-Tekin (ADT) method  yields a conserved charge for each Killing vector field of the background, expressed in terms of an integral over the metric perturbation near spatial infinity. These charges are all zero for the background solution itself; they are otherwise non-zero although convergence of the integrals is not guaranteed. 

Our interest here is in three dimensional (3D) gravity theories;  in particular those that admit an AdS$_3$ vacuum solution, in which case there will also be 
a Ba\~nados-Teitelboim-Zanelli (BTZ) black hole solution \cite{Banados:1992wn}. The BTZ spacetime is parametrized by the dimensionless  constants $(\ell \m,\j)$, 
where $\ell$ is the AdS$_3$ `radius'  and $\m$ is a parameter with dimensions of  mass in units for which $\hbar=1$.  For 3D GR, the ADT method
can be used to show  that $\m$ is the black hole mass $M$, and $\j$ is its angular momentum  $J$ \cite{Bak:1993us,Carlip:1994hq}  (see \cite{Kim:2013zha,Kulkarni:2019nrk} for a more recent discussion).   However, a feature of the ADT method is that it gives the same results  for any two gravitational theories whose field equations become equivalent when linearized about the chosen background solution, and this conflicts with results  obtained by other methods.

This point is most simply illustrated by a comparison of 3D GR with a negative cosmological constant (to allow an AdS$_3$ vacuum) 
to its  `exotic' variant with a  parity-odd  action \cite{Witten:1988hc}.  In this case  the  full field equations (and not just their linearizations) are equivalent, as is most easily
seen from the fact that the action for both can be expressed as a linear combination of two $SL(2;\bR)$ Chern-Simons (CS) actions \cite{Achucarro:1987vz,Witten:1988hc}; after allowing 
for the freedom to rescale the fields and choose the overall sign, there are only two inequivalent linear combinations, corresponding to standard 3D GR and its exotic variant. Using 
a  generalization of the ADT method to CS gravity proposed in \cite{Izquierdo:1994jz}, it was argued in \cite{Townsend:2013ela} that $(\ell M,J) = (\j, \ell \m)$ for exotic gravity.
This exchange of the roles of the two BTZ parameters was found previously in \cite{Carlip:1994hq} for the Carlip-Gegenberg 3D gravity action \cite{Carlip:1991zk}, and a similar role reversal 
occurs  \cite{Afshar:2011qw} for conformal 3D gravity \cite{vanNieuwenhuizen:1985cx,Horne:1988jf}, which also has a parity-odd action. 

These examples suggest that it is important to consider the information contained in the quadratic {\it action} for metric perturbations at infinity, and not just the linearized field equations
that follow from this action.  An $n$th order linearized gravitational field equation defines an $n$th order  partial differential operator but in 3D this operator factorizes (generically) 
into a product of $n$ first-order operators, and the $n$th order quadratic action becomes equivalent to a linear combination of $n$ first-order quadratic actions.  After allowing for 
the freedom to choose the overall sign and redefine fields we are left with a choice of $(n-1)$ relative signs, and hence the possibility of off-shell inequivalences for $n>1$.
The CS gravity example just discussed illustrates this for $n=2$, with the one relative sign distinguishing between the standard and exotic variants of 3D GR.   
A second relative sign  becomes possible for $n=3$,  and this sign is relevant to the comparison between ``Topological Massive Gravity'' (TMG) \cite{Deser:1981wh} and ``Minimal Massive Gravity'' \cite{Bergshoeff:2014pca},
as we explained in a previous work  \cite{Bergshoeff:2018luo}; as we also explained there,  it is essential to take into account these relative signs in any discussion of 
semi-classical unitarity.

Here we discuss similar issues in the context of computations of the values of conserved charges associated to symmetries of a background solution to which other solutions, such as 
BTZ black holes, are asymptotic.  Our  aim is to extend the ADT method, and its CS gravity  generalization,  to  the  ``Chern-Simons-like''  3D gravity 
theories \cite{Hohm:2012vh,Bergshoeff:2014bia,Merbis:2014vja}. These include CS gravity theories as special cases but also the massive 3D gravity theories mentioned
above and many others, such as  ``New Massive Gravity'' (NMG) \cite{Bergshoeff:2009hq}  and the recent ``Exotic Massive Gravity'' (EMG) \cite{Ozkan:2018cxj} which can be viewed as a massive-gravity extension of exotic 3D GR. Because CS-like gravity actions are {\it first-order}, all  relative-sign differences  arising upon linearization are taken into account.

Our main result is a simple and general formula  for conserved charges  as line integrals at spatial infinity.  We use this formula to determine the mass and angular momentum of the 
BTZ black hole as a solution of a variety of 3D gravity theories. In particular, we recover the results mentioned above for the exotic 3D CS gravity, and previous results for TMG and NMG \cite{Bouchareb:2007yx}. We also 
recover the results for MMG reported in \cite{Bergshoeff:2018luo},  and we find the mass and angular momentum of the BTZ black hole solution of EMG.  We should point out that similar results 
for some of these massive gravity theories have been obtained previously  in a series of papers \cite{Setare:2015nla,Setare:2016vhy,Setare:2017wuj,Adami:2017phg},  but many details differ 
from those presented here. 

One of the motivations for this work was the realization that the original ADT method cannot be consistently applied to those 3D massive gravity theories, such as MMG and EMG,  that 
are ``third way consistent'' (in the terminology of \cite{Bergshoeff:2015zga}).  The point here is that for these cases a matter stress tensor is not a consistent source tensor for the metric equation, 
which means that the starting point for the ADT  analysis is not available. There is a consistent source  tensor  \cite{Arvanitakis:2014yja,Ozkan:2018cxj} but it does not reduce to the matter stress tensor even in a linearized limit!  A variant of this difficulty arises even in the simple case of the exotic CS formulation of 3D GR: although it is consistent to add a matter source tensor to the right hand side of the  source-free Einstein equation, this is {\it not} equivalent  to coupling the 3D matter to the dreibein in the usual way (because that would produce a parity-violating equation).  

One advantage of the method used here to determine the asymptotic charges carried by the BTZ black hole is that we start from the most general possible linear coupling
of the one-form fields (which include auxiliary fields for massive 3D gravity) to a generic 2-form source consistent with Noether identities. Ultimately, this source plays no role
in the final formula, as is the case for ADT but now we do not encounter the problem of an inconsistent initial assumption. What we lose is the obvious interpretation of 
asymptotic charges that is provided by the ADT method; for example, it is no longer obvious that the BTZ parameter $\m$ is the mass $M$ of the BTZ black hole, but this was to
to be expected because $M\ne \m$ for many 3D gravity theories, as we have been emphasizing. 

Finally, we compute the central charges of the asymptotic Virasoro $\!\oplus\!$ Virasoro symmetry algebra and discuss the implications of our results for BTZ black hole thermodynamics. 
As we have discussed this topic for MMG in \cite{Bergshoeff:2018luo}, we focus here on the generic Chern-Simons-like model, including EMG, and its 
EGMG generalization \cite{Ozkan:2018cxj}. Our results are both internally consistent and consistent with the discussion in \cite{Townsend:2013ela} for exotic 3D GR, 
but they disagree with some other recent results \cite{Mann:2018vum,Giribet:2019vbj}.

\section{Conserved charges in CS-like gravity}

The generic CS-like action is the integral over a $3$-manifold ${\cal M}$ of a Lagrangian 3-form constructed by exterior multiplication of  $n\ge2$ independent Lorentz 3-vector
valued 1-forms $\{a^r; r=1,2, \dots, n\}$; the Lorentz vector indices are suppressed. The generic Lagrangian 3-form, including a coupling of the one-form fields
to a set of Lorentz 3-vector valued `source' 2-forms $\{{\cal J}_r; r=1,2, \dots n\}$, is 
\begin{equation}
L= \frac12 g_{rs}\,  a^r\cdot da^s + \frac16 f_{rst}\,  a^r\cdot a^s\times a^t - a^r\cdot {\cal J}_r\, , 
\end{equation}
where the exterior products of forms is implicit, and we use standard 3-vector algebra (dot and cross product) for multiplication of Lorentz 3-vectors. The constants 
$g_{rs}$ and $f_{rst}$ can be interpreted as {\it totally symmetric}  tensors on the $n$-dimensional `flavour'  space spanned by $\{a^r; r=1,2, \dots, n\}$; we assume 
that $g_{rs}$ is an invertible metric on this space that we can use to raise or lower `flavour' indices. It is customary to impose some additional conditions but for the moment we proceed with only those just stated. 

The field equations that follow from the above Lagrangian 3-form are 
\begin{equation}\label{eom}
da^r + \frac12 f^r{}_{st} \, a^s\times a^t = {\cal J}^r\, , 
\end{equation}
where we have raised some `flavour' indices with the inverse of $g_{rs}$. 
Let $\{\bar a^r; r=1,2,\dots, n\}$ be a solution of the source free equation; i.e. 
\begin{equation}\label{backfield}
d\bar a^r + \frac12 f^r{}_{st} \, \bar a^s\times \bar a^t =0\, . 
\end{equation}
We may expand about this background solution by writing
\begin{equation}
a^r = \bar a^r + \Delta a^r\, . 
\end{equation}
Substitution into the field equations (\ref{eom}) yields
\begin{equation}\label{backexp}
(\bar D \Delta a)^r + \frac12 f^r{}_{st}\,  \Delta a ^s \times \Delta a^t = {\cal J}^r \, , 
\end{equation}
where,  for any set of Lorentz 3-vector fields $\{V^r\}$ we have
\begin{equation}
(\bar D V)^r := dV^r + f^r{}_{st}\,  \bar a^s \times V^t\, . 
\end{equation}
By defining a ``total source 2-form''
\begin{equation}
{\cal J}_{\rm tot}^r := {\cal J}^r  - \frac12 f^r{}_{st}\,  \Delta a ^s \times \Delta a^t \, , 
\end{equation}
we may rewrite the field equations (\ref{backexp}) in the simple form
\begin{equation}\label{linform}
(\bar D \Delta a)^r = {\cal J}_{\rm tot}^r \, . 
\end{equation}

Now let $\{\xi_r; r=1,\dots,n\}$ be a set of Lorentz-vector scalar fields  (we continue to suppress the Lorentz indices); we may use them to construct the 2-form
\begin{equation}
J =  {\cal J}_{\rm tot}^r \cdot \xi_r \, . 
\end{equation}
A straightforward calculation shows that 
\begin{eqnarray}\label{dJ}
dJ &=& (\bar D {\cal J}_{\rm tot})^r\cdot \xi_r + {\cal J}_{\rm tot}^r \cdot (\bar D \xi)_r \nonumber \\
&=& (\bar D^2 \Delta a)^r \cdot \xi_r + {\cal J}_{\rm tot}^r \cdot (\bar D \xi)_r  \, , 
\end{eqnarray}
where the second equality uses (\ref{linform}). To proceed, we need the following

\begin{itemize}
\item {\bf Lemma}: Let $\{U^r\},\,  \{V^r\}$ be two sets of  Lorentz 3-vectors, either of which may be a set of functions or one-forms on ${\cal M}$. Then 
\begin{equation}
(\bar D^2 V)^r \cdot U_r \equiv - V_r \cdot (\bar D^2 U)^r \, .
\end{equation} 
Proof by calculation:
\begin{eqnarray}
(\bar D^2 V)^r \cdot U_r  &\equiv &  \left\{ f^r{}_{s[t} f^s{}_{u]v} \left[ (\bar a^u\cdot \bar a^t) V^v + 2  \bar a^v  (\bar a^u\cdot V^t) \right]\right\}\cdot U_r \nonumber \\
&\equiv &  f_{rs[t} f^s{}_{u]v} \left[ (\bar a^u\cdot \bar a^t) (V^v\cdot U^r) - 2(V^t\cdot \bar a^u) (\bar a^v\cdot U^r) \right] \nonumber \\
&\equiv&  -f_{rs[t} f^s{}_{u]v} \left[ (\bar a^u\cdot \bar a^t) (V^r\cdot U^v) + 2 (V^r \cdot \bar a^v)   (\bar a^u\cdot U^t) \right] \nonumber \\
&\equiv & - V_r \cdot \left\{ f^r{}_{s[t} f^s{}_{u]v} \left[  (\bar a^u\cdot \bar a^t) U^v + 2 \bar a^v (\bar a^u\cdot U^t) \right] \right\}\nonumber \\
&\equiv &  - V_r \cdot  (\bar D^2 U)^r\, , 
\end{eqnarray}
where the third line uses the antisymmetry of $f_{rs[t} f^s{}_{u]v}$  on the index pair $(v,r)$ and symmetry  under exchange of 
the pairs $(u,t)$ and $(v,r)$.  
\end{itemize}
Using this lemma we have
\begin{equation}
dJ = - \Delta a_r \cdot (\bar D^2 \xi)^r + {\cal J}_{\rm tot}^r \cdot (\bar D \xi)_r  \, . 
\end{equation}
As $(\bar D \xi)^r=0\ \Rightarrow (\bar D^2 \xi)^r=0$, it follows that 
\begin{equation}
(\bar D \xi)^r=0  \quad (r=1,\dots,n) \quad \Rightarrow \quad dJ=0 \, . 
\end{equation}
For the next step we use the field equation (\ref{linform}) to deduce that
\begin{equation}
{\cal J}_{\rm tot}^r \cdot \xi_r  = d \left[ \Delta a^r\xi_r\right] - \Delta a^r \cdot (\bar D \xi)_r\,. 
\end{equation}
It follows that $\bar D\xi=0$ implies $J= d \left[ \Delta a^r\xi_r\right]$, and hence that 
\begin{equation}
\int_\Sigma J = \oint_{\partial\Sigma}  \Delta a^r\cdot \xi_r \, . 
\end{equation}

It remains for us to relate the set of Lorentz-vector scalar fields $\{\xi^r\}$ satisfying $(\bar D\xi)^r=0$ to symmetries of the background. 
This can be done as follows: let 
\begin{equation}
\xi^r = i_\zeta \bar a^r\, , 
\end{equation}
where $i_\zeta$ indicates the contraction with vector field $\zeta$. We have
\begin{eqnarray}
(\bar D\xi)^r &=& d i_\zeta \bar a^r + f^r{}_{st}\,  \bar a^s \times i_\zeta \bar a^t\nonumber \\
&=& \left(d i_\zeta + i_\zeta d\right) \bar a^r - i_\zeta \left(d\bar  a^r +\frac12 f^r{}_{st} \bar a^s\times \bar a^t\right) \nonumber \\
&=& {\cal L}_\zeta \bar a^r \, ,  
\end{eqnarray}
where we have used the background field equations \eqref{backfield} and the formula ${\cal L}_\zeta=di_\zeta + i_\zeta d$ for the Lie derivative of a form field with respect to a vector field $\zeta$. 
We thus see that the background is invariant under a Lie dragging by $\zeta$ when $\bar D\xi^r =0$ for $\xi^r=i_\zeta \bar a^r$, which is the generalization to generic background fields of
the statement that $\zeta$ is a Killing vector field. 
 
 \begin{itemize}
 
 \item {\bf Summary}: For a CS-like gravity theory with a background $\{\bar a^r\}$ such that ${\cal L}_\zeta \bar a^r =0$ ($r=1,\dots,n)$ for vector field $\zeta$, the corresponding conserved 
 charge associated with any configuration $\{ a^r\}$ on a spacelike hypersurface $\Sigma$ that is asymptotic to the background as the boundary circle $\partial\Sigma$ is approached, is 
 \begin{equation}\label{QCSlike}
 Q(\zeta) = \frac{1}{8\pi G} \oint_{\partial\Sigma} \Delta a^r \cdot i_\zeta \bar a^s g_{rs} \, , 
 \end{equation}
 where $\Delta a^r =a^r-\bar a^r$. The normalization will be justified later.

\end{itemize}

This result is a very general one. In order for a 3D gravity interpretation to be possible we must  assume that one linear combination of the  $\{a^r\}$  is the invertible dreibein one-form $e$ from which a Lorentzian metric may be constructed, so the manifold ${\cal M}$ must allow this. In addition, it is customary to also assume that  another linear combination is the dual spin-connection one-form 
$\omega$; the coefficients $g_{rs}$ and $f_{rst}$ are then significantly constrained by the requirement of local Lorentz invariance. We shall impose both these conditions when we turn to applications
of the  formula (\ref{QCSlike})  in the following sections, but neither condition was used in its derivation so it applies more generally. In particular, no assumption of local Lorentz invariance was made, so
(\ref{QCSlike}) will apply to the recent CS-like models of 3D gravity for which this assumption is relaxed \cite{Geiller:2018ain}. 

Here, we choose a basis for the 1-form fields $\{a^r\}$ such that $e$ and $\omega$ are two basis elements, and we insist on local Lorentz invariance. The remaining $(n-2)$ Lorentz vector-valued fields of the basis (for $n>2$) will be assumed to be auxiliary in the sense that they are determined algebraically in terms of $e$ and $\omega$ by the full set of field equations.  We shall also  insist on explicit closed form expressions for these auxiliary fields (without resort to an infinite-series expansion) but we reconsider this condition in our final Discussion section.

\section{BTZ black hole charges}\label{sec:3}

We  now aim to apply the formula (\ref{QCSlike}) to determine the mass and angular momentum of the BTZ spacetime in the context of various 3D gravity theories, starting with 
the CS gravity theories.  The dreibein components are
\begin{equation}\label{BTZdreibein}
e^0 = N(r) dt\, , \qquad e^1 = r\left(d\varphi + N^\varphi dt\right)  \, , \qquad e^2= \frac{dr}{N(r)} \, , 
\end{equation}
where 
\begin{subequations}
\begin{eqnarray}\label{Ns}
N^2 &=& \frac{(r^2-r_+^2)(r^2-r_-^2)}{\ell^2 r^2} = \frac{r^2}{\ell^2} - 8 G \, \m + \left(\frac{4 G\, \j}{r}\right)^2 \, , \\
N^\varphi &=& \frac{r_+r_-}{\ell r^2}  = \frac{4 G \, \j}{r^2} \, .
\end{eqnarray}
\end{subequations}
The (outer and inner) horizon radii $r_+$ and $r_-$ are related to the parameters $\m$ and $\j$ by 
\begin{equation}
\ell \m= \frac{r_+^2 + r_-^2}{8 \ell G} \, , \qquad \j = \frac{r_+ r_-}{4 \ell G}\, . 
\end{equation}
The zero-torsion condition determines the dual Lorentz connection one-form:
\begin{eqnarray}\label{BTZspincon}
\omega^0 = - Nd\varphi  \, , \qquad \omega^1 = - \frac{4 G\, \j}{r}d\varphi - \frac{r}{\ell^2} dt \, , \qquad \omega^2 = \frac{4 G\, \j}{r^2 N} dr\, . 
\end{eqnarray}

We shall take the background to be the ``black hole vacuum'' for which $\m=0$ and $\j=0$, so that 
\begin{equation}\label{ebar}
\bar e^0 =  \frac{r}{\ell} dt\, , \qquad \bar e^1 = rd\varphi\, ,   \qquad \bar e^2 =  \frac{\ell}{r} dr \, ,
\end{equation}
and 
\begin{equation}\label{omegabar}
\bar \omega^0 = - \frac{r}{\ell} d\varphi\, , \qquad \bar\omega^1 =  - \frac{r}{\ell^2} dt  \, , \qquad \bar\omega^2 = 0\, , 
\end{equation}
and hence 
\begin{eqnarray}\label{Deltae}
\Delta e^0 &=& \left[N- \frac{r}{\ell}\right] dt = -\frac{4 G \ell \m}{r} dt + \dots \nonumber \,, \\
\Delta e^1 &=& \frac{4 G\, \j}{r} dt\, , \\
\Delta e^2 &=& \left[N^{-1} - \frac{\ell}{r}\right] dr =  \frac{4 G \ell^3 \m}{r^3} dr + \dots  \, ,\nonumber 
\end{eqnarray}
and 
\begin{eqnarray}\label{Deltaomega}
\Delta\omega^0 &=& - \left(N-\frac{r}{\ell}\right) d\varphi =  \frac{4 G \ell \m}{r} d\varphi + \dots \, , \nonumber \\
\Delta\omega^1 &=& - \frac{4 G\, \j}{r} d\varphi \, , \\
\Delta\omega^2 &=& \frac{\ell \j}{2r^2 N} dr\nonumber =  \frac{4 G \ell \j}{ r^3}dr + \ldots  \, ,
\end{eqnarray}
where omitted terms are subleading in the $r\to\infty$ limit.

Now we consider in turn 3D GR and its exotic variant, taking $\Sigma$ to be a surface of constant $t$.  The isometries of the BTZ black hole vacuum 
correspond to the two Killing vector fields $\partial_t$ and $\partial_\phi$, and our principal interest here is to determine the relation between 
the corresponding conserved charges for the BTZ black hole in terms of its parameters $(\m,\j)$.

\subsection{Standard 3D GR}

The Lagrangian 3-form for 3D gravity with cosmological constant $\Lambda=-1/\ell^2$ is 
\begin{equation}
(8\pi G) L_{\rm GR} =  - e\cdot R(\omega) - \frac{1}{6\ell^2} e\cdot e\times e \, ,  
\end{equation} 
where $G$ is the 3D Newton constant (which has dimensions of inverse mass).  After addition of the exact 3-form $\frac12d[\omega\cdot e]$, the  right hand side takes the CS-like form with
\begin{equation}
g_{e\omega}= -1 \, , \quad  f_{e\omega\omega}= - 1  \,,  \quad  f_{eee}= - \frac{1}{\ell^2} \, . 
\end{equation}
Application of the formula (\ref{QCSlike}) yields
\begin{equation}\label{QGR}
(8\pi G)Q_{\rm GR}(\zeta) = -  \oint_{\partial\Sigma} \left[ \Delta e \cdot \bar\omega_\mu + \Delta\omega \cdot \bar e_\mu \right]\zeta^\mu \, . 
\end{equation}

Now we consider in turn the two (dimensionless) Killing vector fields $\ell\partial_t$ and $\partial_\varphi$ of the black-hole vacuum background.
Using eqs. (\ref{ebar} - \ref{Deltaomega}) we find that
\begin{subequations}
\begin{eqnarray}
Q_{\rm GR}(\ell \partial_t) &=& - \frac{\ell}{8\pi G} \oint_{\partial\Sigma} \left[\Delta e^1 \bar\omega^1_t - \Delta\omega^0 \bar e^0_t\right] = 
\frac{\ell}{8\pi G}  \oint 4 G\, \m d\varphi  = \ell \m \, , \\ 
Q_{\rm GR}(\partial_\varphi) &=& - \frac{1}{8\pi G} \oint_{\partial\Sigma} \left[ -\Delta e^0 \bar\omega^0_\varphi + \Delta\omega^1 \bar e^1_\varphi \right] = \frac{1}{8\pi G} \oint 4 G\, \j d\varphi = \j \, .
\end{eqnarray}
\end{subequations}
This agrees with standard results if we make the identification 
\begin{equation}
\ell M= Q(\ell \partial_t) \, , \qquad J= Q(\partial_\varphi) \, , 
\end{equation}
and this fact justifies our choice of normalization in (\ref{QCSlike}). In other words, with this normalization and the above identification of the charges with the 
BTZ black hole mass and angular momentum we have $(M,J)=(\m,\j)$ for 3D GR. 

\subsection{Exotic 3D GR} 

The Lagrangian 3-form for ``Exotic Gravity'' is 
\begin{equation}\label{EGLag}
(8\pi G) L_{\rm EG} = \ell\,  L_{\rm LCS}  + \frac{1}{2\ell} \, e \cdot T(\omega) \, , 
\end{equation}
where $L_{LCS}$ is the Lorentz-Chern-Simons  3-form for $\omega$,  and $T(\omega)$ is its torsion 2-form:
\begin{equation}
L_{\rm LCS} = \frac12 \left[ \omega \cdot d \omega + \frac13 \omega \cdot \omega \times \omega\right] \,, \qquad T(\omega) = de + \omega\times e \equiv D(\omega) e\, . 
\end{equation}
In this case the right hand side of (\ref{EGLag}) is of CS-like form with 
\begin{equation}
g_{ee}=\frac{1}{\ell}  , \qquad g_{\omega\omega} = \ell \, , \qquad  f_{ee\omega} = \frac{1}{\ell}\, , \qquad  f_{\omega\omega\omega}= \ell\, ,
\end{equation}
and hence
\begin{equation}
(8\pi G) Q_{\rm EG}(\zeta) = \oint_{\partial\Sigma} \left[\ell \Delta\omega \cdot \bar\omega_\mu + \frac{1}{\ell} \Delta e\cdot \bar e_\mu \right]\zeta^\mu \, .
\end{equation}

It is useful to notice here (and for calculations to follow) that 
\begin{equation}\label{lemma} 
 \oint_{\partial\Sigma} \Delta e\cdot \bar e_\mu \zeta^\mu =0\,  \qquad ({\rm for}\ \  \zeta= \ell\partial_t \ \ {\rm and}\ \ \zeta = \partial_\varphi)\, , 
 \end{equation}
and hence 
\begin{equation}\label{QEG} 
 (8\pi G) Q_{\rm EG}(\zeta) = \ell \oint_{\partial\Sigma} \left[\Delta\omega \cdot \bar\omega_\mu \right]\zeta^\mu \, . 
 \end{equation}
On substitution for $\zeta$, this formula shows that 
\begin{subequations}
\begin{eqnarray}
Q_{\rm EG} (\ell \partial_t) &=& \frac{\ell^2}{8\pi G} \oint_{\partial\Sigma} \left[ \Delta\omega^1  \bar\omega^1_t  \right] = 
\frac{\ell^2}{8\pi G} \oint_{\partial\Sigma} \frac{4G\, \j}{\ell^2} \, d\varphi  =  \j \, , \\
 Q_{\rm EG} (\partial_\varphi) &=& \frac{\ell}{8\pi G} \oint_{\partial\Sigma} \left[- \Delta\omega^0  \bar\omega^0_\varphi \right] = \frac{\ell}{8\pi G}\oint 4G\, \m \, d\varphi =  \ell \m \, .
\end{eqnarray}
\end{subequations}
Notice that the charges for exotic 3D gravity are exchanged with respect those of standard 3D GR,  in agreement with \cite{Townsend:2013ela}:
\begin{equation}
Q_{\rm EG}(\ell \partial_t) = Q_{\rm GR}(\partial_\varphi) \,, \qquad   Q_{\rm EG}(\partial_\varphi) = Q_{\rm GR}(\ell \partial_t) \, . 
\end{equation}

\subsection{Conformal 3D gravity}

A CS-like action for 3D conformal gravity is \cite{Hohm:2012vh}
\begin{equation}
L_{\rm CG} = k\left[ L_{\rm LCS} + h\cdot T(\omega) \right]\, , 
\end{equation}
for arbitrary dimensionless constant $k$.  This reduces to the Van Nieuwenhuizen action \cite{vanNieuwenhuizen:1985cx} on solving the zero-torsion constraint imposed by $h$, and it is 
a partially gauge-fixed version of the  Horne-Witten CS action \cite{Horne:1988jf}. From this CS-like action we may read off the $g_{rs}$ coefficients:
\begin{equation}
g_{\omega\omega} =k\, , \qquad g_{eh} =k \, .
\end{equation}

The field equations allow for an AdS$_3$ solution of arbitrary `radius' $\ell$ and hence for a BTZ black hole solution. In either case one finds that 
\begin{equation}
h= \frac{1}{2\ell^2} \, e\, , 
\end{equation}
and hence that 
\begin{equation}
\bar h = \frac{1}{2\ell^2} \, \bar e\, , \qquad \Delta h = \frac{1}{2\ell^2} \, \Delta e\, . 
\end{equation}
Taking (\ref{lemma}) into account, one finds that 
\begin{equation}
Q_{\rm CG}(\zeta) = \frac{k}{\ell} Q_{\rm EG}(\zeta)\, . 
\end{equation}
This confirms the role reversal of the BTZ parameters  for the BTZ black hole as a solution of 3D conformal gravity as compared with 3D GR  \cite{Afshar:2011qw}: 
whereas one might have expected $Q_{\rm CG}(\zeta)$ to be a factor times $Q_{\rm GR} (\zeta)$, it is instead a factor times $Q_{\rm EG}(\zeta)$.  As for exotic 3D gravity, this exchange of roles
is a consequence of having a parity-odd action for parity-preserving field equations.

\section{Massive 3D gravity theories} \label{sec:Massive}

For the class of CS-like theories considered in this paper, the Lagrangian 3-form is constructed from a set of $n$ Lorentz-vector valued one-form fields
$\{a^r; r=1,2, \dots\}$, one of which  is the dreibein $e$ (assumed to be invertible) and another is the Lorentz-dual spin connection $\omega$ (which is 
required to ensure local Lorentz invariance). The other $(n-2)$ fields are assumed to be auxiliary, in the sense that the full set of equations of motion can be 
used to solve for them algebraically, in closed form in terms of the dreibein\footnote{In this sense $\omega$ is also auxiliary but we do not include it among the
``auxiliary'' fields because of its special status.}.   This has the following implication: {\it for any locally maximally-symmetric solution of the full equations
all auxiliary fields will be proportional to $e$}. 

In this paper, $n\le4$, so it will be sufficient to consider the $n=4$ case for which there are two auxiliary fields; let us call them $(h,f)$. As the BTZ black hole 
solution is locally maximally symmetric (because it is locally equivalent to AdS$_3$) we have
\begin{equation}\label{prop}
h = c_h e\, , \qquad f= c_f  e \, ,
\end{equation}
for this solution, where $(c_h,c_f)$  are (model-dependent) constants, which implies that
\begin{equation}
\bar h = c_h \bar e\, , \qquad \bar f= c_f \bar e\, ; \qquad \Delta h = c_h \Delta e\, , \qquad \Delta f= c_f \Delta e\, . 
\end{equation}
This allows us to simplify the formula (\ref{QCSlike}) in applications of it to the 3D massive gravity theories of interest here (for which $n=3,4$). This formula reduces to 
\begin{equation}\label{Qeff}
Q(\zeta) =  \oint_{\partial\Sigma}\left\{ \left(\Delta e\cdot \bar \omega_\mu + \Delta\omega \cdot \bar e_\mu \right) g^{\rm eff}_{e\omega} + \Delta \omega \cdot \bar \omega_\mu\,  g^{\rm eff}_{\omega\omega}
+ \Delta e\cdot \bar e_\mu \, g^{\rm eff}_{ee} \right\} \zeta^\mu\, , 
\end{equation}
where 
\begin{eqnarray}\label{geff}
g^{\rm eff}_{e\omega} &=& g_{e\omega} + c_h g_{h\omega} + c_f g_{f\omega}\, , \nonumber\\
g^{\rm eff}_{\omega\omega} &=& g_{\omega\omega} \, , \\
g^{\rm eff}_{ee} &=& g_{ee} + 2 c_h g_{he} + 2 c_f g_{fe} + c_h^2 g_{hh} +  2 c_h c_f g_{hf} + c_f^2 g_{ff} \, . \nonumber
\end{eqnarray}
The  $g^{\rm eff}_{ee}$ coefficient is irrelevant to the final result for the charges as a consequence of (\ref{lemma}). However, it will be relevant later
when we compute the central charges of the asymptotic symmetry algebra; we shall then use the fact that the restrictions on the $g_{rs}$ and $f_{rst}$ coefficients
that follow from the assumption of an AdS$_3$ vacuum solution are such that 
\begin{equation}\label{gee}
g_{ee}^{\rm eff} = \frac{1}{\ell^2} g_{\omega\omega}^{\rm eff}\,. 
\end{equation}
To see this consider the background field equation $g_{rs} d\bar a^s + \frac12 f_{rst} \bar a^s \times \bar a^t = 0$ for $r = \omega$. 
Making use of \eqref{prop} and local Lorentz invariance (which imposes $f_{\omega r s} = g_{r s}$) this equation becomes
\begin{equation}
g_{\omega e}^{\rm eff} \bar D \bar e + g_{\omega\omega}^{\rm eff} R(\bar \omega) = - \frac{1}{2} g_{ee}^{\rm eff} \bar e \times \bar e \,.
\end{equation}
Demanding that AdS$_3$ is a solution with vanishing torsion will then impose \eqref{gee}. One could interpret this equation as fixing the cosmological constant in terms of the Chern-Simons-like coupling constants.

Finally, using the expressions (\ref{QGR}) and (\ref{QEG}), we deduce that 
\begin{equation}\label{simplef}
Q(\zeta) =  - g^{\rm eff}_{e\omega}\,  Q_{\rm GR}(\zeta) + \ell^{-1} g^{\rm eff}_{\omega\omega} Q_{\rm EG}(\zeta) \, . 
\end{equation}
This formula greatly simplifies the calculations to follow for various 3D massive gravity theories. 

\subsection{TMG, NMG and GMG} 

The ``General Massive Gravity'' (GMG) model introduced in \cite{Bergshoeff:2009hq} propagates a pair of spin-2 modes with arbitrary (non-zero) masses $m_\pm$; parity is violated 
when $m_+\ne m_-$. The special case for which $m_+=m_-=m$ is the parity preserving ``New Massive Gravity'', also introduced in \cite{Bergshoeff:2009hq}; it propagates a parity doublet
of spin-2 modes of mass $m$. The special case for which $m_+ \to \infty$ for finite $m_-=\mu$ yields the ``Topologically Massive Gravity'' model of \cite{Deser:1981wh}; this propagates
a single spin-2 mode of mass $\mu$. Finally, TMG reduces to 3D GR in the $\mu\to \infty$ limit. 

The  Lagrangian 3-form for GMG is \cite{Hohm:2012vh}
\begin{eqnarray}
(8\pi G) L_{GMG} &=&  - \sigma e \cdot R(\omega) + \frac16 \Lambda_0\,  e \cdot e \times e + h \cdot T(\omega)  + \frac{1}{\mu} L_{LCS}  \nonumber \\
&& \qquad +\, \frac{1}{m^2} f\cdot R(\omega) - \frac{1}{2m^2} e\cdot f\times f \, . 
\end{eqnarray}
By the addition of an exact 3-form, this can be put into CS-like form with
\begin{equation}
g_{e\omega} = -\sigma \, , \qquad g_{eh} = 1 \, , \qquad g_{f\omega} = - \frac{1}{m^2}\, , \qquad g_{\omega\omega} = \frac{1}{\mu} \, . 
\end{equation}
In the AdS$_3$ vacuum with $\Lambda= -1/\ell^2$, and for the BTZ solution, we have
\begin{equation}
\ell^2 \Lambda_0 = -\sigma + \frac{1}{4(\ell m)^2} \, , 
\end{equation}
and $(h,f) =(c_h,c_f) e$ with 
\begin{equation}
c_h = \frac{1}{2\mu\ell^2} \, , \qquad c_f = \frac{1}{2\ell^2} \, . 
\end{equation}
Applying the formula (\ref{geff}) we find that 
\begin{equation}\label{geffGMG}
g^{\rm eff}_{e\omega} = - \left(\sigma + \frac{1}{2(\ell m)^2} \right) \, , \qquad g^{\rm eff}_{\omega\omega} = \frac{1}{\mu}\, , 
\end{equation}
and hence
\begin{equation}
Q_{GMG}(\zeta) = \left(\sigma + \frac{1}{2(\ell m)^2} \right)Q_{GR}(\zeta) + \frac{1}{\ell\mu} Q_{EG}(\zeta)\, . 
\end{equation}
This gives us
\begin{equation}
\ell M_{\rm GMG} = \left(\sigma + \frac{1}{2(\ell m)^2} \right)\ell\m + \frac{\j}{\ell\mu}\, , \qquad J_{\rm GMG}= \left(\sigma + \frac{1}{2(\ell m)^2} \right)\j + \frac{\m}{\mu} \, .
\end{equation}

From this result we get the result for NMG by taking the $\mu \to\infty$ limit and the result for TMG by taking the  taking the $m\to\infty$ limit:
\begin{subequations}
\begin{align}
\ell M_{\rm NMG} &= \left(\sigma + \frac{1}{2(\ell m)^2} \right)\ell\m \, , &  J_{\rm NMG} & = \left(\sigma + \frac{1}{2(\ell m)^2} \right)\j \, , \\
\ell M_{\rm TMG} &= \sigma \ell\m + \frac{\j}{\ell\mu}\, , &  J_{\rm TMG} & = \sigma \j + \frac{\m}{\mu}\, . 
\end{align}
\end{subequations}

\subsection{MMG} 

The deformation of TMG to MMG \cite{Bergshoeff:2014pca} consists of adding an interaction term to the TMG Lagrangian with parameter $\alpha$: 
\begin{equation}
	(8\pi G) L_{\rm MMG} = (8\pi G) L_{\rm TMG} + \frac{\alpha}{2} e \cdot h \times h \,.
\end{equation}
One might expect that the new interaction term would have no effect on  the TMG  result because the formula \eqref{QCSlike} appears to depend only on the (unchanged) coefficients $g_{rs}$ of the kinetic terms. 
However, there is an implicit dependence on the interaction coefficients $f_{rst}$ because this affects the background solution; for the case in hand one finds that $\omega$ is not torsion-free  when $\alpha\ne0$. 
The torsion-free  connection is 
\begin{equation}
\Omega = \omega + \alpha h\,.
\end{equation}
In terms of this new connection the MMG Lagrangian 3-form is \cite{Bergshoeff:2018luo} 
\begin{align}\label{Lmmg2}
(8\pi G) L_{\rm MMG} = & - \sigma e \cdot R(\Omega) + \frac{\Lambda_0}{6} e \cdot e \times e + (1+ \alpha\sigma) h \cdot T(\Omega) - \frac{\alpha}{2}(1+\alpha \sigma) e\cdot h \times h \nonumber \\
 &  + \frac{1}{\mu} L_{\rm LCS} (\Omega) - \frac{\alpha}{\mu} h \cdot \left( R(\Omega) - \frac{\alpha}{2} D(\Omega) h + \frac{\alpha^2}{6} h\times h \right)\,, 
\end{align}
where $D(\Omega)$ denotes the Lorentz covariant derivative with respect to the connection $\Omega$, and (as explained in \cite{Bergshoeff:2014pca}) the definition of MMG includes the restriction 
$(1+ \alpha \sigma) \neq 0$.  For this Lagrangian 3-form, the $g_{rs}$ coefficients are
\begin{equation}
g_{e\Omega} = - \sigma\, , \qquad g_{\Omega\Omega} = \frac{1}{\mu} \, , 
\end{equation}
and 
\begin{equation}
g_{eh} = (1+ \alpha\sigma) \, , \qquad g_{hh}= \frac{\alpha^2}{\mu}\, , \qquad g_{h\Omega}= - \frac{\alpha}{\mu}\, . 
\end{equation}

For the BTZ solution,  $e$ and $\Omega$ are given by the 3D GR (and TMG) expressions for $e$ and $\omega$, and $h= c_h e$ with 
\begin{equation}
c_h = \mu C \, , 
\end{equation}
where the dimensionless constant $C$ is determined, along with $\Lambda_0$, by the equations
\begin{equation}\label{MMGvacrel}
C = \frac{(1-\alpha\ell^2 \Lambda_0)}{2[(1+\alpha\sigma)\mu\ell]^2} \, , \qquad \Lambda_0 = - \frac{\sigma}{\ell^2} + \alpha \left[(1+\alpha\sigma)\mu C\right]^2\, . 
\end{equation}In this case $h$ is the only auxiliary field, so 
\begin{subequations}\label{geffMMG}
	\begin{eqnarray}
	g^{\rm eff}_{e\Omega} &=& g_{e\Omega} + c_h \, g_{h\Omega} = - \left[ \sigma + \alpha C\right] \, ,  \\
	g^{\rm eff}_{\Omega\Omega}  &=& g_{\Omega\Omega} = \frac{1}{\mu}\, . 
	\end{eqnarray}
\end{subequations}
As $\Omega$ is the same torsion-free connection as used previously in the TMG case, the expressions we need for $\bar\Omega$ and $\Delta\Omega$ 
are exactly the same as those used in the TMG case for $\bar\omega$ and $\Delta\omega$. The formula (\ref{simplef}) therefore continues to apply, and in this case it
tells us that 
\begin{equation}
Q_{\rm MMG} (\zeta) = \left(\sigma +  \alpha C\right) Q_{\rm GR} (\zeta)   +  \frac{1}{\mu\ell} Q_{\rm EG} \, . 
\end{equation}
As expected, we recover the TMG result upon setting $\alpha=0$. 

Applying this result for the two Killing vector fields of the BTZ black hole vacuum, we deduce that 
\begin{equation}
\ell M_{\rm MMG} = (\sigma + \alpha C)\ell \m  +  \frac{\j}{\ell \mu} \, ,  \qquad J_{\rm MMG} =  (\sigma + \alpha C)\j  + \frac{\m}{\mu} \, , 
\end{equation}
which agrees with  \cite{Bergshoeff:2018luo}.

\subsection{EMG and EGMG}

Exotic massive gravity (EMG) \cite{Ozkan:2018cxj} is the exotic 3D gravity version of NMG  \cite{Bergshoeff:2009hq}; both propagate a parity doublet of spin-2 modes, but the EMG equations are found from 
an odd-parity CS-like Lagrangian 3-form. A generalization that leads to a parity-violating metric field equation was also found in \cite{Ozkan:2018cxj} and called there ``Exotic Generalized Massive Gravity''  (EGMG); its 
Lagrangian 3-form is 
\begin{eqnarray}\label{EMGaction}
(8\pi G) L_{\rm EGMG} &=& -\frac{\ell}{m^2}  \bigg[ f \cdot R(\omega) + \frac{1}{6m^4} f \cdot f \times f - \frac{1}{2m^2} f \cdot D(\omega) f + \frac{\nu}{2} f \cdot e \times e 
\nonumber\\
&& \qquad - \ m^2 h \cdot T(\omega) + (\nu - m^2) L_{LCS}(\omega)  + \frac{ \nu m^4}{3 \mu} e \cdot e \times e \bigg]\,, 
\end{eqnarray}
where\footnote{We require $\nu\ne0$ here although the EGMG metric equation has a well-defined $\nu\to0$ limit \cite{Ozkan:2018cxj}.} 
\begin{equation}
\nu = \frac{1}{\ell^2} - \frac{m^4}{\mu^2} \,.
\end{equation}
The EMG Lagrangian 3-form is obtained by taking the $\mu\to\infty$ limit. 

We may read off from (\ref{EMGaction}) the non-zero $g_{rs}$  coefficients that we will  need to apply the formula \eqref{QCSlike}:
\begin{equation}
g_{eh} = \ell \, , \qquad g_{ff} = \frac{\ell}{m^4} \, , \qquad g_{\omega\omega} = \ell\left(1 - \frac{\nu}{m^2}\right) \, , \qquad  g_{f\omega} = - \frac{\ell}{m^2} \, . 
\end{equation}
The BTZ black hole is a solution of EMG when the components of the dreibein and spin connection are given as in \eqref{BTZdreibein} and \eqref{BTZspincon}, respectively, and 
\begin{equation}\label{auxfield}
h = c_h e\,, \qquad  f= c_f e \, , 
\end{equation}
for  constants 
\begin{equation}
c_f = - \frac{m^4}{\mu}\,, \qquad c_h = \frac{1}{2\ell^2}(1-\frac{1}{\ell^2 m^2})(1- \frac{\ell^2 m^4}{\mu^2})\, . 
\end{equation}
From this we learn that 
\begin{equation}\label{geffEGMG}
g^{\rm eff}_{e\omega} = \frac{\ell m^2}{\mu}\, , \qquad g^{\rm eff}_{\omega\omega} = \ell\left(1+ \frac{m^2}{\mu^2} - \frac{1}{(\ell m^2)} \right)\, , 
\end{equation} 
and hence that 
\begin{equation}\label{QEGMG}
Q_{\rm EGMG}(\zeta) = - \frac{\ell m^2}{\mu} Q_{\rm GR}(\zeta) + \left(1+ \frac{m^2}{\mu^2} - \frac{1}{(\ell m^2)} \right) Q_{\rm EG}(\zeta) \, .
\end{equation}

By taking the $\mu\to\infty$ limit we get the corresponding result for EMG:
\begin{equation}
Q_{\rm EMG}(\zeta) = \left(1- \frac{1}{(\ell m)^2}\right) Q_{\rm EG}(\zeta) \, .
\end{equation}

This result here differs from the ADT charges computed using the linearized field equations in \cite{Mann:2018vum}. As a check on our result, we observe that 
\begin{equation}
\lim_{m^2\to\infty} Q_{EMG}(\zeta) =  Q_{EG}(\zeta) \, .
\end{equation}
This could have been anticipated from the fact that
\begin{equation}
(8\pi G) \lim_{m^2\to\infty} L_{\rm EMG} = \ell h\cdot T(\omega) + \ell L_{\rm LCS}(\omega) \, , 
\end{equation}
which is the Lagrangian 3-form for exotic 3D GR after a re-interpretation of $2\ell^2 h$ as a new dreibein.


\section{Central charges for CS-like theories} \label{sec:assym}

Under the assumptions listed in the previous sections, it becomes possible to derive a generic formula for the central charges of the putative holographic duals to the various CS-like theories of gravity discussed here. This derivation rests on the realization that the formula \eqref{QCSlike} still applies for asymptotic diffeomorphisms, which become true Killing symmetries only at the spatial boundary where $r \to \infty$. The treatment of asymptotic symmetries in Chern-Simons-like theories of gravity was discussed in \cite{Grumiller:2017otl} and reviewed in \cite{Bergshoeff:2018luo}, we will not repeat this analysis here in full detail. We only need that generic solutions with asymptotically AdS$_3$ boundary conditions are described by their Fefferman-Graham expansion, which is finite in three dimensions and yields the Ba\~nados metrics given in \eqref{banados} below.  We then find the asymptotic diffeomorphisms preserving the Ba\~nados metric in all CS-like theories with an AdS$_3$ solution for which the auxiliary fields satisfy $(h,f) =  (c_h, c_f)e$, and we use the algebra of asymptotic charges to compute the boundary central charges.

\subsection{Asymptotic charges for Ba\~nados solutions}

Ba\~nados metrics \cite{Banados:1998gg} parameterize the phase space of locally asymptotically AdS$_3$ solutions. They are given in terms of two arbitrary state-dependent functions $\cL^\pm(x^\pm)$ as
\begin{align}\label{banados}
ds^2 = & \; dr^2 - \ell^2\left(e^{r/\ell} dx^+ - e^{-r/\ell} \cL^-(x^- ) dx^-\right) \left(e^{r/\ell} dx^- - e^{-r/\ell} \cL^+(x^+ ) dx^+\right)\, ,
\end{align}
where $x^\pm = t \pm \varphi$. We are interested in studying transformations of this background which correspond to true symmetries of the background only asymptotically. In other words, we wish to impose
\begin{equation}\label{assym}
(D \xi)^r = d\xi^r + f^r{}_{st} a^s \times \xi^t = \delta_\xi a^r \, ,
\end{equation}
where $\delta_\xi a^r \to 0 $ only at the asymptotic boundary $r \to \infty$. The sub-leading components of \eqref{assym} will determine how the state-dependent functions $\cL^\pm$ transform under asymptotic symmetry transformations. 

To proceed we first parameterize the Ba\~nados metrics \eqref{banados} by the following dreibein
\begin{subequations}\label{eq:dreibein}
	\begin{align}
	e^0 & = \frac{\ell}{2} \left( 2 e^{r/\ell} -  e^{-r/\ell} (\cL^+ + \cL^-)\right)dt - \frac{\ell}{2} e^{-r/\ell}(\cL^+ - \cL^-) d\varphi\,, \label{e0}  \\
	e^1 & =   \frac{\ell}{2} e^{-r/\ell}(\cL^+ - \cL^-) dt  + \frac{\ell}{2} \left( 2 e^{r/\ell} +  e^{-r/\ell} (\cL^+ + \cL^-)\right) d\varphi \, , \label{e1}\\
	e^2 & = dr   \,. \label{e2}
	\end{align}
\end{subequations}
After solving the torsion constraint $de+\omega \times e=0$ for the spin-connection we find 
\begin{subequations}\label{eq:spincon}
	\begin{align}
	\omega^0 & = \frac{1}{2} \left( - 2 e^{r/\ell} +  e^{-r/\ell} (\cL^+ + \cL^-)\right)d\varphi + \frac{1}{2} e^{-r/\ell}(\cL^+ - \cL^-)dt   \,, \label{omega0}  \\
	\omega^1 & = - \frac{1}{2} e^{-r/\ell}(\cL^+ - \cL^-) d\varphi - \frac{1}{2} \left( 2 e^{r/\ell} +  e^{-r/\ell} (\cL^+ + \cL^-)\right) dt    \,, \label{omega1} \\
	\omega^2 & = 0\,.  \label{omega2}
	\end{align}
\end{subequations}
We wish to find the gauge parameters $\xi^r$ which preserve the form of $a^r$ up to a transformation of the state-dependent functions $\cL^{\pm}$. In general this may be a non-trivial task for the generic CS-like theory, however under our working assumptions the problem simplifies. 

Let us first use that the auxiliary fields $(h,f)$ are proportional to the dreibein. This implies for gauge parameters $\xi^r$ corresponding to asymptotic diffeomorphisms that also $(\xi^h, \xi^f) = (c_h, c_f) \xi^e$. Hence the only two independent components of $\xi^r$ that we have to solve for are $\xi^e $ and $\xi^\omega$, corresponding to (possibly  linear combinations of) an asymptotic diffeomorphism and a local Lorentz transformation.

Then we use that local Lorentz invariance imposes restrictions on the structure constants of the CS-like theory involving $\omega$. To be precise, for local Lorentz invariance of the action (and corresponding field equations) we need 
\begin{equation}
f^r{}_{\omega r} = 1 \quad \& \quad  f^r{}_{\omega s} = 0  \quad \text{for } \, r \neq s\,.
\end{equation}
The first of these conditions can be understood as the statement that every derivative $d$ needs to be accompanied by a spin-connection $\omega$, and the second condition states that all spin connections arise in this way. 

Under these assumptions equation \eqref{assym} for $r = (e, \omega)$ becomes 
\begin{subequations}\label{deltaeo}
\begin{align}
d \xi^e + \omega \times \xi^e + e \times \xi^\omega + c_1 e \times \xi^{e} = \delta_\xi e\, ,\\
d \xi^\omega + \omega \times \xi^\omega + c_2 e \times \xi^e = \delta_\xi \omega \, ,
\end{align}
\end{subequations}
with
\begin{subequations}
\begin{align}
c_1 & = f^e{}_{ee} + 2 c_h f^e{}_{eh} + 2 c_f f^e{}_{ef} + c_h^2 f^{e}{}_{hh} + c_f^2 f^e{}_{ff} + 2c_h c_f f^e{}_{hf} \, ,  \\
c_2 & = f^\omega{}_{ee} + 2 c_h f^\omega{}_{eh} + 2 c_f f^\omega{}_{ef} + c_h^2 f^{\omega}{}_{hh} + c_f^2 f^\omega{}_{ff} + 2c_h c_f f^\omega{}_{hf} \, .
\end{align}
\end{subequations}
At the same time, solving the field equations for the Ba\~nados metrics with vanishing torsion and cosmological constant $\Lambda = - 1/\ell^2$ tells us that 
\begin{equation}\label{c12}
c_1 = 0 \,, \qquad c_2 = \frac{1}{\ell^2} \,.
\end{equation}
Likewise, the conditions on the structure constants and $c_h, c_f$ that arise from solving the field equations for the Ba\~nados solutions guarantee that \eqref{assym} is satisfied for $r = (h,f)$. 

What remains to be done is to solve \eqref{deltaeo} with \eqref{c12}. The solution can be parameterized by two arbitrary functions $f^\pm(x^\pm)$ and reads
\begin{align}\label{xie}
\xi^e  = &  \, \frac{\ell}{2} e^{-r/\ell} \left( f^+ ( e^{2r/\ell} - \cL^+ ) + f^- (e^{2r/\ell} - \cL^-) + \frac12(f^{+}{}'' +  f^{-}{}'' ) \right)T^0   \\
& + \, \frac{\ell}{2} e^{-r/\ell} \left( f^+ ( e^{2r/\ell} + \cL^+ ) - f^- (e^{2r/\ell} + \cL^-) - \frac12(f^{+}{}'' -  f^{-}{}'' ) \right)T^1  \nonumber \\
& - \frac12(f^+{}' + f^-{}' ) T^2 \, ,\nonumber 
\end{align}
and
\begin{align}\label{xiomega}
\xi^\omega  = &  \, - \frac{\ell}{2} e^{-r/\ell} \left( f^+  ( e^{2r/\ell} - \cL^+ ) - f^- (e^{2r/\ell} - \cL^-) + \frac12(f^{+}{}'' -  f^{-}{}'' ) \right)T^0   \\
& - \, \frac{\ell}{2} e^{-r/\ell} \left( f^+ ( e^{2r/\ell} + \cL^+ ) + f^- (e^{2r/\ell} + \cL^-) - \frac12(f^{+}{}'' +  f^{-}{}'' ) \right)T^1  \nonumber \\
& + \frac12(f^+{}' - f^-{}' ) T^2\,. \nonumber
\end{align}
Here $T^a \; (a=0,1,2) $ are the generators of $SO(1,2)$.
The transformation of the state-dependent functions $\cL^{\pm}$ is then given by 
\begin{align}\label{eq:cfttrans}
\delta_\xi \cL^\pm = f^\pm \cL^\pm{}' + 2 f^\pm{}' \cL^{\pm} - \frac12 f^\pm{}'''\,.
\end{align}
These are the transformation properties of the left and right moving stress tensors of a conformal field theory and $\xi^{r}$ (for $r = (e, h,f)$) encode the usual Brown-Henneaux asymptotic Killing vectors $\zeta^\mu$ by $\xi^{r} = a_{\mu}^r \zeta^\mu$.\footnote{This expression holds for $r = \omega$ only up to a local Lorentz transformation which is sub-leading towards the boundary and does not contribute to the asymptotic charges.}

Now that we have found the asymptotic transformations \eqref{assym} with parameters $\xi^r$ we apply our formula for the asymptotic charges to obtain
\begin{align}
Q[\xi^r] & = \frac{1}{8\pi G} \oint_{\partial \Sigma}  \Delta a^r \cdot \xi^s g_{rs} \nonumber \\
& = \frac{1}{8\pi G} \oint _{\partial \Sigma} \left( ( \Delta e \cdot \xi^\omega + \Delta \omega \cdot \xi^e) g_{e\omega}^{\rm eff} +  \left( \frac{\Delta e \cdot \xi^e}{\ell^2}  + \Delta \omega \cdot \xi^\omega \right) g_{\omega\omega}^{\rm eff} \right) \, .
\end{align}
The effective coefficients $g^{\rm eff}_{rs}$ are given in \eqref{geff} and we have used the relation \eqref{gee}. Using the Ba\~nados solutions \eqref{eq:dreibein} and \eqref{eq:spincon} with $\cL^\pm = 0$ as background values for $\bar e$ and $\bar \omega$ we have
\begin{subequations}\label{eq:deltae}
	\begin{align}
	\Delta e^0 & = - \frac{\ell}{2} e^{-r/\ell} \left( (\cL^+ + \cL^-) dt + (\cL^+ - \cL^-) d\varphi \right)\,, \\
	\Delta e^1 & =   \frac{\ell}{2} e^{-r/\ell} \left( (\cL^+ - \cL^-) dt + (\cL^+ + \cL^-) d\varphi \right)\,, \\
	\Delta e^2 & = 0 \,,
	\end{align}
\end{subequations}
and
\begin{subequations}\label{eq:deltaom}
	\begin{align}
	\Delta \omega^0 & = \frac{1}{2}  e^{-r/\ell} \left(  (\cL^+ + \cL^-) d\varphi + (\cL^+ - \cL^-) dt \right)   \,,  \\
	\Delta \omega^1 & = - \frac{1}{2} e^{-r/\ell} \left( (\cL^+ - \cL^-) d\varphi + (\cL^+ + \cL^-) dt \right)   \,,  \\
	\Delta \omega^2 & = 0\, .
	\end{align}
\end{subequations}
Together with \eqref{xie} and \eqref{xiomega} this implies that 
\begin{equation}\label{Qbanados}
Q = \frac{\ell}{8\pi G} \oint_{\partial \Sigma}  \left[ - g_{e\omega}^{\rm eff} (\cL^+ f^+ + \cL^- f^-)  + \frac{1}{\ell} g_{\omega\omega}^{\rm eff}  (\cL^+ f^+ - \cL^- f^-)\right] \, .
\end{equation}

Ba\~nados geometries \eqref{banados} with constant $ \cL^{\pm} = \frac{2G}{\ell} (\ell \m \pm \j)$ describe the BTZ black hole geometry. The Killing vector $\partial_t$ corresponds to taking $f^{\pm} = 1$ and $\partial_\varphi$ corresponds to $f^\pm = \pm 1$, which allows us to recover the result \eqref{simplef} from this formula as well. For a generic asymptotic symmetry transformation parameterized by the functions $f^{\pm} (x^\pm)$ these charges become generators of the 2D conformal algebra, with central charges depending on the coefficients of the CS-like model, as we will now show.


\subsection{Central charges}\label{sec:cc}
Let us first specialise to the case of 3D general relativity (GR) and exotic gravity (EG). The result for GR is
\begin{equation}
Q_{\rm GR}[\zeta_\mu] = \frac{\ell}{8\pi G} \oint d \varphi \left( f^+(x^+) \cL^+(x^+) + f^-(x^-) \cL^-(x^-) \right)\,,
\end{equation}
whereas exotic 3D gravity gives:
\begin{equation}
Q_{\rm EG}[\zeta_\mu] = \frac{\ell}{8\pi G} \oint d \varphi \left( f^+(x^+) \cL^+(x^+) - f^-(x^-) \cL^-(x^-) \right)\,.
\end{equation}
The results differ only by a sign but this difference has major consequences. Using the transformation property \eqref{eq:cfttrans} we see that the Poisson brackets of the charges $\{ Q[f], Q[g]\} = \delta_f Q[g]$ span two copies of the Virasoro algebra in both cases, but the two central charges for exotic gravity have opposite sign: 
\begin{equation}\label{EGcc}
c_{\rm GR}^\pm =  \frac{3\ell}{2G}\,, \qquad c_{\rm EG}^{\pm} = \pm \frac{3\ell}{2G}\,.
\end{equation}
Since the transformation property \eqref{eq:cfttrans} is universal for asymptotically AdS$_3$ spacetimes in the sense that it does not depend on the specifics of the CS-like model, the algebra of charges still consists of two copies of the Virasoro algebra in the generic case with charges \eqref{Qbanados}. In contrast, the values of the central charges {\it do}  depend on the specifics of the CS-like model,  and they can be written as linear combinations of the GR and EG central charges: 
\begin{align}\label{ccgeneric}
c^\pm & = - g_{e\omega}^{\rm eff} c^\pm_{\rm GR} +  \frac{1}{\ell} g_{\omega\omega}^{\rm eff} c^\pm_{\rm EG}  = \left( - g_{e\omega}^{\rm eff} \pm \frac{1}{\ell} g_{\omega\omega}^{\rm eff} \right) \frac{3\ell}{2G}\,.
\end{align}
We shall now verify this formula for the known cases discussed in the last section and then compute the central charges of EGMG.

\subsubsection{GMG}
For GMG we have computed $g^{\rm eff}_{e\omega} $ and $g^{\rm eff}_{\omega\omega}$ in \eqref{geffGMG}.
Application of the formula \eqref{ccgeneric} gives
\begin{equation}
c^\pm_{\rm GMG} = \left(\sigma + \frac{1}{2(\ell m)^2} \pm \frac{1}{ \mu \ell} \right) \frac{3\ell}{2G} \,,
\end{equation}
which indeed corresponds to the central charge of GMG as reported in \cite{Bergshoeff:2009hq}.

\subsubsection{MMG}
In the case of MMG, we first change variables to $\Omega = \omega + \alpha h$ such that $\Omega$ is the torsionless spin-connection. We then apply the formula \eqref{ccgeneric} with $\omega$ replaced by $\Omega$ together with \eqref{geffMMG}. The result is
\begin{equation}
c^{\pm}_{\rm MMG} = \left( \sigma + \alpha C \pm \frac{1}{\mu\ell} \right) \frac{3\ell}{2G} \,,
\end{equation}
in agreement with \cite{Bergshoeff:2014pca,Bergshoeff:2018luo}\,.

\subsubsection{EGMG}
We now turn to the computation of the central charges of EGMG. Using the effective coefficients \eqref{geffEGMG} our formula \eqref{ccgeneric} straightforwardly leads to the following expression for the EGMG central charges\footnote{These expressions are proportional to the coefficients $a_\pm$ found from the quadratic action  in \cite{Ozkan:2018cxj}, in agreement with the claim made 
there that $a_\pm \propto c^\pm$, but the constant of proportionality is negative here because of  slightly different conventions.}
\begin{equation}\label{ccEGMG}
	c^\pm_{\rm EGMG} = \left( - \frac{\ell m^2}{\mu} \pm \left( 1 + \frac{m^2}{\mu^2} - \frac{1}{(\ell m)^2} \right) \right) \frac{3\ell}{2G} \, .
\end{equation}
The limit of $\mu \to \infty$ gives the EMG central charges
\begin{equation}
c^\pm_{\rm EMG} =  \pm \left( 1  - \frac{1}{(\ell m)^2} \right) \frac{3\ell}{2G} \, .
\end{equation}
The limit $m \to \infty$ then reproduces the EG central charges, not those of 3D GR.

Note that $c^\pm = 0$ now has two solutions, corresponding to two critical values of $\mu$ where one of the boundary central charges vanishes, these are
\begin{equation}
\mu_{\rm crit,1} = \pm \ell m^2 \,, \qquad \mu_{\rm crit,2} = \pm \frac{\ell m^2}{\ell^2 m^2 - 1}\,.
\end{equation} 
It would be interesting to see whether the logarithmic solutions to EMG studied in \cite{Mann:2018vum,Giribet:2019vbj} at $\mu_{\rm crit,2}$ are also solutions at $\mu_{\rm crit,1}$.

The EGMG central charges \eqref{ccEGMG} differ from those reported in \cite{Giribet:2019vbj}. The authors of that paper derived the central charges by integrating the first law of black hole thermodynamics using the ADT-charges, computed in the metric formulation in \cite{Mann:2018vum}, and then assuming the validity of Cardy's formula. In the next section we will see that integrating the first law with the EGMG black hole charges \eqref{QEGMG} will give an entropy consistent with Cardy's formula when the central charges are given by \eqref{ccEGMG}. In fact, we will show that Cardy's formula holds for any CS-like theory satisfying our assumptions.

\section{Black hole thermodynamics}
\label{sec:thermo}

The first law of black hole thermodynamics states that a black hole of mass $M$ and angular momentum $J$ satisfies the first law of thermodynamics
\begin{equation}\label{firstlaw}
	dM - \Omega dJ = T dS \,,
\end{equation}
where $\Omega$ is the angular velocity, $T$ is the Hawking temperature of the black hole and $S$ the black hole (Bekenstein-Hawking) entropy. 
The mass $M$ and angular momentum $J$ are extensive properties and will in general depend on the specific theory under consideration, as will the entropy.
In general relativity the entropy corresponds to the area of the outer horizon at $r=r_+$ over $4G$ in Planck units, but for exotic gravity the entropy is proportional to the area of the inner horizon at $r=r_-$ \cite{Townsend:2013ela}.

From the results of section \ref{sec:Massive} we see that for the CS-like theories considered in this paper we have
\begin{subequations}
\begin{align}
	\ell M 	& = Q(\ell \partial_t) = - g^{\rm eff}_{e\omega}\,  \ell \m + g^{\rm eff}_{\omega\omega} \, \frac{\j}{\ell} \, , \\
	J 		& = Q(\partial_\varphi) = -g^{\rm eff}_{e\omega}\,  \j + g^{\rm eff}_{\omega\omega}\,  \ell\m \, .
\end{align}
\end{subequations}

The parameters   $(\m, \j)$ of the BTZ black hole spacetime,  as well as the Hawking temperature and angular velocity of the black hole (which are intensive variables), do not depend on the specific theory and can be expressed solely in terms of the horizon radii of the BTZ solution:
\begin{subequations}\label{btzrelations}
\begin{align}
\m & = \frac{r_+^2 + r_-^2}{8 \ell^2 G}\,, & \j & = \frac{r_+ r_-}{4\ell G} \,, \\
T & = \frac{r_+^2 - r_-^2}{2\pi \ell^2  r_+} \,, & \Omega  & = \frac{r_-}{\ell r_+} \,.
\end{align} 
\end{subequations}
Using these relations one can integrate the first law \eqref{firstlaw} and find the entropy of the BTZ black hole in the generic Chern-Simons-like theory of gravity. The result is
\begin{equation}\label{genericS}
S = - g^{\rm eff}_{e\omega}\,  \frac{2\pi r_+}{4G} + \frac{g^{\rm eff}_{\omega\omega} }{\ell} \, \frac{2\pi r_-}{4G}\,.
\end{equation}
It is now straightforward to reproduce the known results for the various theories under consideration:
\begin{subequations}\label{Entropies}
\begin{align}
S_{\rm GR} & = \frac{\pi r_+}{2 G} \,, \qquad S_{\rm EG} = \frac{\pi r_-}{2 G} \,, \\
S_{\rm GMG} & = \left(\sigma + \frac{1}{2(\ell m)^2} \right) \frac{\pi r_+}{2 G} + \frac{1}{\mu\ell} \frac{\pi r_-}{2 G}\,, \\
S_{\rm MMG} & = \left(\sigma + \alpha C \right) \frac{\pi r_+}{2 G} + \frac{1}{\mu\ell} \frac{\pi r_-}{2 G}\,.
\end{align}
\end{subequations}
When applying our formula \eqref{genericS} to EGMG we find the result
\begin{equation}
	S_{\rm EGMG} = - \frac{\ell m^2}{\mu}  \frac{\pi r_+}{2 G} + \left(1+ \frac{m^2}{\mu^2} - \frac{1}{(\ell m)^2} \right) \frac{\pi r_-}{2 G}\,.
\end{equation}
In the limit $\mu \to \infty$ this reduces to the EMG entropy 
\begin{equation}
S_{\rm EMG} =  \left(1 - \frac{1}{(\ell m)^2} \right) \frac{\pi r_-}{2 G}\,,
\end{equation}
which further reduces to the EG entropy when $m \to \infty$.


\subsection{Cardy formula}
Now we wish to verify whether Cardy's formula holds for the entropy of the generic CS-like models. 
Using the result for the central charge \eqref{ccgeneric}, the entropy of BTZ black holes \eqref{genericS} can be accounted for by the Cardy formula in the form
\begin{equation}\label{CardyT}
S = \frac{\ell \pi^2}{3}(c^+ T^+ + c^- T^-)\, , \qquad \left(T^\pm = \frac{r_+ \pm r_-}{2\pi \ell^2} \right)
\end{equation}
where $T^\pm$ are the left/right BTZ temperatures. 

We wish to express the entropy in the more familiar form of the Cardy formula,
\begin{equation}\label{Cardysqrt}
S_{\rm Cardy} =  2\pi \sqrt{\frac{c_+ }{6} \Delta^+} + 2\pi \sqrt{\frac{c_- }{6} \Delta^-}\, ,
\end{equation}
where $\Delta^{\pm}$ are the eigenvalues of the Virasoro generators $L_0^{\pm}$. These are the zero modes of \eqref{Qbanados} for Ba\~nados geometries describing the BTZ black hole, i.e. metrics \eqref{banados} with $\cL^{\pm} = \frac{2G}{\ell} (\ell \m \pm \j)$. Using this and the relations \eqref{btzrelations} we find
\begin{equation}
\Delta^{\pm} = \frac{\left[- g_{e\omega}^{\rm eff} \pm g_{\omega\omega}^{\rm eff} /\ell\right] }{16 \ell G} \, (r_+ \pm r_-)^2 \,.
\end{equation}
The Cardy formula \eqref{Cardysqrt} then yields
\begin{equation}\label{Sintermediate}
S_{\rm Cardy} =  \frac{\pi}{6\ell} \left( (r_+ + r_-) |c^+| + (r_+ - r_-) |c^-| \right) \,,
\end{equation}
which agrees with \eqref{genericS} only when both $c^+> 0$ and $c^->0$.  
Even though any unitary CFT meets this assumption, we do not wish to assume $c^{\pm} > 0$ here because $c^- <0$. for exotic gravity.   It was already noted in \cite{Townsend:2013ela} that in this case Cardy's formula \eqref{Cardysqrt} should come with a minus sign in front of the second term; from \eqref{Sintermediate} we see that in general we should take
\begin{equation}\label{Cardysqrtgen}
	S = {\rm sign}(c^+) 2\pi \sqrt{\frac{c_+}{6}\Delta^+} + {\rm sign}(c^-) 2\pi \sqrt{\frac{c_-}{6} \Delta^-} \,.
\end{equation}
This is because a negative sign for $c^+$ (or $c^-$) can be understood as taking all left-movers (or right-movers) to have energy $E<0$ with a Hamiltonian bounded from above. In order to maintain the standard thermodynamical relations this should change the partition function as $Z_L \to Z_L^{-1}$ (or $Z_R \to Z_R^{-1}$) and hence a minus sign in front of the relevant term in the Cardy formula appears \cite{Townsend:2013ela}.


\section{Discussion}

The Chern-Simons-like formulation of massive 3D gravity theories is a simple, but surprisingly powerful,  generalization of the Chern-Simons formulation of 3D General Relativity. 
It greatly simplifies an analysis of the properties of  models like ``Topologically Massive Gravity'' (TMG) and ``New Massive Gravity''  (NMG),  which were first found by other means,  
and it leads  naturally to the ``Minimal Massive Gravity''  (MMG) and ``Exotic Massive Gravity''  (EMG) generalizations that could not have been  easily found in any other way. It 
also leads to a simple determination of the requirements for perturbative unitarity and, in those cases for which there is an AdS$_3$  vacuum, a determination of the central charges 
of the asymptotic 2D conformal algebra. 

Here we have shown that the formalism also leads, by an adaptation of the Abbott-Deser-Tekin method,  to a simple integral formula for asymptotic charges associated to 
solutions that are asymptotic to an AdS$_3$ vacuum. Application to the mass $M$ and angular momentum $J$ of the BTZ black hole solution yields a general formula
for each of these quantities as a linear combination of the two parameters $(\m,\j)$ of the BTZ spacetime with coefficients that are linear or quadratic functions of the parameters  of
the CS-like action.   The  3D CS gravity theories are special cases to which this formula also applies, and for 3D GR  it yields the expected result that $M=\m$ and $J=\j$. 
Applied to the ``exotic'' variant of 3D GR (with parity-odd action)  the formula yields the result that $M= \j/\ell$ and $J=\ell \m$ in agreement with earlier results. 

In their CS formulations,  it is only a choice of sign that distinguishes 3D GR from its exotic variant, and this sign is irrelevant to the field equations. 
The original ADT method cannot distinguish between 3D GR and its exotic variant because it makes use {\it only} of the field equations. 
This is not obviously a problem but it is certainly a limitation in the context of the AdS/CFT correspondence because this is a conjectured equivalence between 
the partition function of a CFT as a function of sources for CFT operators and (in a particular limit) a  path-integral over the action of a  classical gravity theory with 
corresponding asymptotic boundary conditions. Different actions can be expected to correspond to different CFTs. 

It is actually only the {\it linearized} equations (about the AdS background) that are used in the ADT method, so this method cannot distinguish 
between any two 3D gravity models with the same linearized equations. An example of such a pair is TMG and MMG and, as we emphasised in a
previous paper \cite{Bergshoeff:2018luo}, the fact that the quadratic action for these two CS-like models is first-order leads to the possibility of an off-shell inequivalence 
arising from a sign difference that has no effect on the linearized field equations.  The variant of the ADT method presented here for CS-like theories
takes account of this different sign and thus gives results for MMG  that differ from those of TMG (but agree with our own earlier results). 

Another pair of 3D gravity models with equivalent linearized equations is NMG and EMG. Here the off-shell inequivalence of the respective quadratic actions 
is obvious because one is parity even and the other is parity odd. We have applied our formula to both NMG (finding agreement with earlier results) and EMG,
for which we find results that are in disagreement with those found recently in \cite{Mann:2018vum,Giribet:2019vbj}. The main reason for this disagreement is that the authors of \cite{Mann:2018vum,Giribet:2019vbj} apply the ADT-method to the linearized EMG field equations in the {\it metric formulation} but their stress tensor source  is not the linearized limit of any 
consistent source tensor for the full EMG equations \cite{Ozkan:2018cxj}. By coupling a generic source tensor to all Chern-Simons-like one-form fields we find here results which are both internally consistent and consistent with black hole thermodynamics and Cardy's formula.

Although our formula \eqref{QCSlike} for the asymptotic charges  is valid for any Chern-Simons-like theory, the simplifications for the asymptotic charges of the BTZ black hole \eqref{simplef} and the central charges \eqref{ccgeneric} apply only to theories in which all Lorentz-vector one-form fields  other than the dreibein and spin connection are auxiliary. Specifically, it must be possible
to solve the CS-like field equations for any additional Lorentz-vector 1-forms in terms of $e$ and $\omega$,  and hence in terms of $e$ since we also assume (possibly after a field redefinition that preserves the CS-like form of the action)  that the CS-like equations imply that $\omega$ is torsion-free; these auxiliary fields will then be proportional to $e$ in the BTZ solution of the full CS-like field equations. This condition is met by all the 3D gravity models discussed so far\footnote{Excepting the Carlip-Gegenberg 3D gravity model \cite{Carlip:1991zk}) and possibly the Geiller-Noui models in which local Lorentz invariance is not imposed \cite{Geiller:2018ain}.} and  by many other CS-like theories, examples of which may be  found in  \cite{Afshar:2014ffa} and the recent \cite{Ozkan:2019iga,Afshar:2019npk}. 

We also assumed that the expressions for auxiliary fields in terms of $e$ are closed form expressions rather than infinite-series expansions; this assumption was made in 
order to avoid issues such as the convergence of infinite series, and because it is satisfied for most of the best-known 3D gravity theories, but there is no obvious reason our results 
should not apply if this condition is relaxed.   There is thus reason to suppose that the results obtained here will be applicable to Zwei-Dreibein Gravity \cite{Bergshoeff:2013xma},  and generalizations thereof \cite{Afshar:2014dta}.   

Our simplified central charge formula \eqref{ccgeneric} will clearly not apply if the background solution is warped AdS$_3$ \cite{Nutku:1993eb,Gurses:1994bjn}  because the asymptotic symmetry algebra is then different, and neither will our  simplified formula \eqref{simplef} for black hole charges apply, because the black hole solution is different \cite{Moussa:2003fc,Anninos:2008fx}. 
However, we would still expect our formula \eqref{QCSlike} to apply.

\subsection*{Acknowledgements}
Some of the work for this paper was done during the 2019 Amsterdam Summer Workshop on String Theory and we thank the organisers for their hospitality. 
WM is supported by the ERC Advanced Grant {\it High-Spin-Grav}  and by FNRS-Belgium (convention FRFC PDR T.1025.14 and convention IISN 4.4503.15). 
PKT is is partially supported by the STFC consolidated grant ST/L000385/1.



\providecommand{\href}[2]{#2}\begingroup\raggedright\endgroup

\end{document}